\documentclass[a4paper]{panl}
\usepackage{cite}
\usepackage{wrapfig}
\usepackage{graphicx}
\graphicspath{ {./images/} }
\DeclareGraphicsExtensions{.pdf,.png}
\usepackage{amssymb}
\usepackage{amsfonts}
\usepackage{color}
\usepackage{amsmath}
\usepackage{longtable}
\usepackage{rotating}
\usepackage{lscape}
\usepackage{epsfig}
\usepackage{multirow}
\usepackage[hidelinks]{hyperref}
\usepackage[justification=centering]{caption}
\usepackage{lineno}
\usepackage{footnote}
\usepackage{subcaption}
\usepackage{cleveref}
\makesavenoteenv{tabular}
\makesavenoteenv{table}
\usepackage{xcolor}
\hypersetup{
	colorlinks,
	linkcolor={red!80!black},
	citecolor={green!80!black},
	urlcolor={blue!80!black}
}
\originalTeX
\begin{document}

\title{On the possibility to study antiproton production at the SPD detector at NICA collider for dark matter search in astrophysical experiments. }
\maketitle
\authors{A.\,Guskov$^{a,}$\footnote{E-mail: alexey.guskov@cern.ch},
R.\,El-Kholy$^{b,}$\footnote{E-mail: relkholy@sci.cu.edu.eg}}
\from{$^{a}$\,Joint Institute for Nuclear Research (JINR), Dubna, Russia, 141980}
\vspace{-3mm}
\from{$^{b}$\,Astronomy Department, Faculty of Science, Cairo University, Giza, Egypt, 12613}

\begin{abstract}
Dark matter is an important component of the Standard model of cosmology but its nature is still unknown. One of the most common explanations is that dark matter consists of Weakly Interacting Massive Particles (WIMPs), supposed to be cold thermal relics of the Big Bang and to build up the galactic dark halos. Indirect search of dark matter could be performed via the study of an anomalous antiproton component in cosmic rays originating from possible annihilation of dark matter pairs in the galactic halo, on top of the standard astrophysical production. The measurements performed by the AMS-02 and PAMELA spectrometers have shown that limited knowledge of antiproton-production cross sections in $pp$, $pD$, $pHe$ and $HeHe$ collisions is one of the main uncertainties of background subtraction. The planned SPD experiment at the NICA collider could provide a precision measurement of antiproton yield in wide kinematic range in $pp$ and $pD$ collisions at the energy scale from the threshold to $\sqrt{s}=26$ GeV/$c$.
\end{abstract}
\vspace*{6pt}


\label{sec:intro}
\section*{\textsc{Introduction}}~

Evidence indicating that Dark Matter (DM) is the dominant matter constituent in the Universe has been compiling. This DM is thought to enclose the large structures of the Universe, warping the spacetime fabric, and thus shaping the field of gravity in the Universe. Even though its origin and nature are still unknown, one of the most common theories is that DM consists of Weakly Interacting Massive Particles (WIMPs), which are assumed to be thermal relics of the Big Bang and to build up the galactic dark halos.

Indirect search for DM is based on the possibility of detecting anomalies in rare components of Cosmic Rays (CRs), namely, antiparticles (especially, antiprotons), produced via annihilation or decay of DM particles. This idea motivated powerful new experiments; among them are the AMS-02 and PAMELA spectrometers. However, in order to be able to detect any exotic anomalous components that may exist, it is crucial to first measure with absolute precision the production of the same components from conventional astrophysical sources.

The secondary antiprotons background is produced in proton-nucleus, nucleus-proton, and nucleus-nucleus collisions of CRs on the Interstellar Medium (ISM). However, nuclei heavier than helium have a marginal role in the secondary antiproton-production. The largest contributions are from reactions that involve protons and helium (i.e. $pp$, $pHe$, $He~p$, and $HeHe$). Even though experimental data on antiprotons are limited, AMS-02 have measured both the antiproton flux and the antiprotons to protons ratio ($\bar{p}/p$). These measurements have reached an unprecedented accuracy of a few percents---exceeding even the latest PAMELA results of antiprotons\cite{adriani2013}---and extended the range of absolute rigidity to 450 GeV \cite{ams2016}, showing that the $\bar{p}/p$ reaches a plateau above $\sim 60$ GeV with respect to absolute rigidity\cite{ams2016}.

\begin{figure}[htb]
	\centering
	\includegraphics[width=\textwidth]{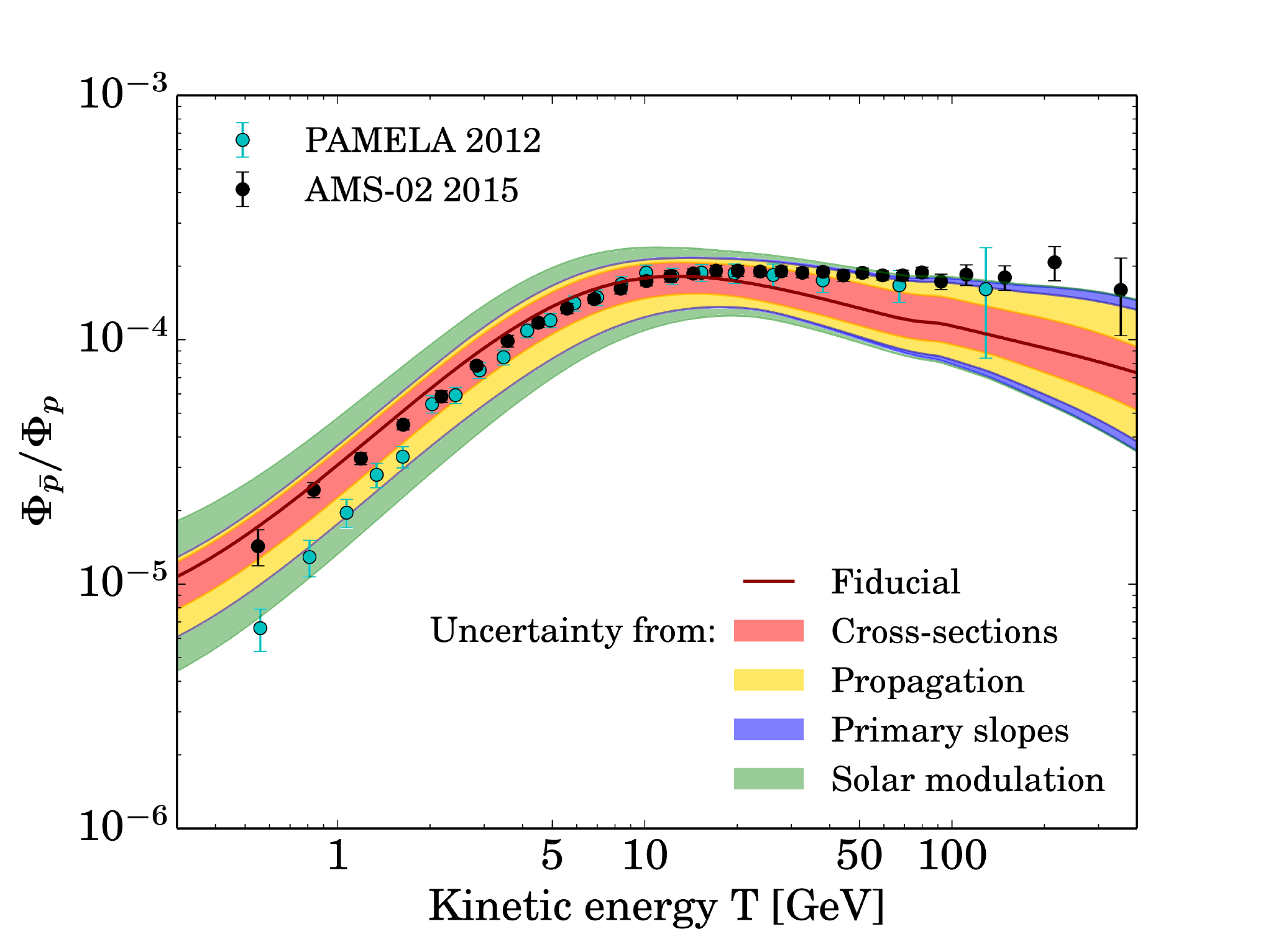}
	\caption{The combined total uncertainty on the predicted secondary $\bar{p}/p$, superimposed to the PAMELA and the AMS-02 data. Each envelope accounts for uncertainties enclosed in it; such that the green band is the total band of all four sources of uncertainty\cite{uncertainties}.}
	\label{uncertainties}
\end{figure}

However, even with such high precision, any predictions made about the secondary antiproton component or any exotic origins that might exist would be strongly affected by several sources of uncertainty. These sources are: i) uncertainties on the primary spectra slopes at high energies, ii) uncertainties on antiproton-production cross sections, especially on reactions involving helium where there are almost \emph{no data}, iii) uncertainties on the propagation parameters that determine diffusion and convection in the local galactic environment, which in turn determine the contributions of any hypothetical primary, or exotic antiproton fluxes that reach the Earth, and iv) uncertainties on the Solar modulation which involves several parameters that depend on the Solar activity at the time of observation\cite{uncertainties}. While the last could be minimized by conservative variation of these parameters, and the first and third can be relatively reduced through the AMS-02 measurements, new measurements have to be performed to reduce uncertainties regarding production cross sections which vary from about 20\% to 50\% at most, depending on the energy\cite{uncertainties}. All these sources of uncertainties are depicted in Fig. \ref{uncertainties}\cite{uncertainties} as colored envelopes around a fiducial curve with reference values for the different sources of uncertainties, and are superimposed to the PAMELA\cite{adriani2013} and AMS-02\cite{ams2016} data of the $\bar{p}/p$.

In addition to the scarcity of antiproton-production data, most of the existing datasets lack an important correction. Some of the antiprotons produced in collision experiments and in CRs are a result of the decay of intermediate hyperons; namely, $\bar{\Lambda}$ and $\bar{\Sigma}^{-}$. While newer experiments apply a feed-down to correct the data, older ones had not been taking that into account. Thus, relevant corrections have to be applied to old datasets. And even though the ratio of antiprotons produced via hyperon decay is not large at low energies, it gains significance as the energy increases\cite{Winkler:2017xor}. Moreover, in older studies of CRs, it was assumed that antineutrons and antiprotons were equally produced. So, antiprotons from decaying antineutrons were assumed to be equal to promptly-produced antiprotons. And due to the lack of data on antineutron production, estimations were only made through theoretical arguments. However, it was recently remarked\cite{kappl2014,mauro2014} that the isospin of the colliding particles might favor the production of one particle or the other.

It is therefore evident that accurate measurements of antiproton-production cross sections are fundamentally important for coming to any conclusions about exotic signals from DM annihilation or decay in the AMS-02 data. Moreover, these measurements would need to cover a wide range of energies for the proton beam from 10 GeV to 6 TeV, and a pseudorapidity ($\eta$) range from 2 to about 8, or, in terms of the CM frame, a transverse momentum ($p_{T}$) range from 0.04 to 2 GeV for produced antiprotons and a radial-scaling variable ($x_{R}$) range from 0.02 to 0.7, in order to keep up with the AMS-02 antiproton energy range\cite{PhysRevD.96.043007}. Such measurements would allow phenomenological models of the CRs antiproton flux from standard astrophysical sources to reach the same level of accuracy as the AMS-02; hence, allowing comparison of analogous predictions to come to a conclusion about the existence of an antiproton component from DM annihilation or decay.

In this work, we are going to briefly review the production of antiprotons in CRs collisions on the ISM. Then we will summarize the existing data on antiproton production. Finally, we will discuss future measurements planned and the possibility of studying antiproton production at the Spin Physics Detector (SPD) at the Nuclotron-based Ion Collider fAсility (NICA).

\label{sec:pbar_production}
\section{\textsc{Antiproton Production in Hadronic Collisions}}~

Passing through space, particles of CRs interact with matter of the ISM.
The abundances of the light isotopes in CRs  in the energy range of about a few GeV and in the ISM are presented in Tab. \ref{tab:abundance}. The estimated relative contribution of interaction of the components of CRs and the ISM into the inclusive antiproton production according to\cite{uncertainties} is listed in Tab. \ref{tab:contributions}. The $pD$ and $p^3$He collisions are not mentioned in\cite{uncertainties}. Their contribution is estimated  based on the abundance of deuterium and $^3$He in CRs and the assumption that the antiproton-production cross sections are 1/2 and 3/4, respectively, of the $p^4$He cross section.

\begin{table}[htbp]
\begin{center}
  \caption{Abundances $n_i$ of the light isotopes in CRs\cite{denolfo2000, abundances2001} and in the ISM\cite{wilson1994}.}
  \label{tab:abundance}
  \begin{tabular}{ccc}
  \hline\hline
 Nuclei & Cosmic Rays & Interstellar Medium \\
    \hline
  $p$         &  0.844 &  0.911\\
  $D$         &   0.029  & $1.6\times10^{-5}$\\
  $^3$He    &  0.027    & $2\times10^{-5}$\\
  $^4$He    &  0.10  & 0.089 \\
  \hline\hline
  \end{tabular}
\end{center}
\end{table}

\begin{table}[htbp]
\begin{center}
  \caption{Relative contributions from interaction of the components of CRs and the ISM into the inclusive antiproton production\cite{uncertainties}.}
  \label{tab:contributions}
  \begin{tabular}{cc}
  \hline\hline
Collision & $\bar{p}$ yield normalized to  \\
               & the $\bar{p}$ yield in $pp$ collisions \\
    \hline
  $pp$         &  1 \\
  $p$$^4$He         &   0.4 \\
  $^4$He$^4$He    &  0.014  \\
  $pD$     &  0.06 \\
  $p$$^3$He  &  0.08\\
 
  \hline\hline
  \end{tabular}
\end{center}
\end{table}

There are several distinct mechanisms through which antiprotons are produced. The most dominant one is direct production through proton-proton scattering ($p + p \longrightarrow \bar{p} + X$). However, a considerable contribution to antiproton production comes from decay of anti-baryons. In general, due to the conservation of baryon number and the antiproton being the only stable one, any anti-baryon will eventually decay to an antiproton. In particular, decays of long-lived intermediate states, namely the $ \bar{\Lambda} $ and $ \bar{\Sigma}^{-} $ hyperons and antineutrons, to antiprotons can be a convenient tool for estimation of the number of those anti-baryons\cite{kappl2014}. While the $\bar{\Lambda}$ hyperon decays into an antiproton and a pion $(\bar{\Lambda}^{0} \longrightarrow \bar{p} + \pi^{+})$ with a branching factor of $63.9 \pm 0.5 \%$\cite{pdg2012}, there is a less dominant contribution\cite{kappl2014} from the decay of  $\bar{\Sigma}^{-}$ hyperons into an antiproton and a pion $(\bar{\Sigma}^{-} \longrightarrow \bar{p} + \pi^{0})$. That decay mode has a branching factor of $51.57 \pm 0.30 \%$\cite{pdg2012}.

Other sources include the beta decay of antineutrons into antiprotons $(\bar{n} \longrightarrow \bar{p} + e^{+} + \nu_{e})$ with a branching factor of approximately $100 \%$. Generally speaking, antiprotons are also produced in proton-nucleus and nucleus-nucleus scattering $(p + X \longrightarrow p + \bar{p} + p + X)$, even though there is not enough data for these interactions\cite{kappl2014}.

In order to detect any exotic origins of antiproton production (e.g. DM annihilation), we first need to optimize our models of other standard sources of antiprotons. However, there still is a discrepancy of accuracy between theoretical predictions and corresponding experiments, which hinders interpretation of experimental data of antiproton production. The antiproton-production cross sections in CRs inelastic interactions are considered one of the main sources of uncertainty\cite{mauro2014}. Moreover, precision data for differential cross sections of antiproton production in proton-nucleus, nucleus-proton, and nucleus-nucleus collisions are almost non-existent. While nuclei heavier than helium do not play a large role in the process, reactions involving helium represent a considerable fraction of antiproton production. This stresses the need for data if we were to maximize the use of upcoming data from AMS-02. However, for the case of $pp$ collisions, experimental data are more abundant \cite{brahms2007, na49-2010, PhysRev.137.B962, Allaby:1970jt
, CAPILUPPI1974189, GUETTLER197677, PhysRevLett.39.1173, PhysRevD.19.764}.

\begin{figure}[!htb]
	\centering
	\begin{subfigure}[t]{0.495\textwidth}
		\centering
		\includegraphics[width=\linewidth]{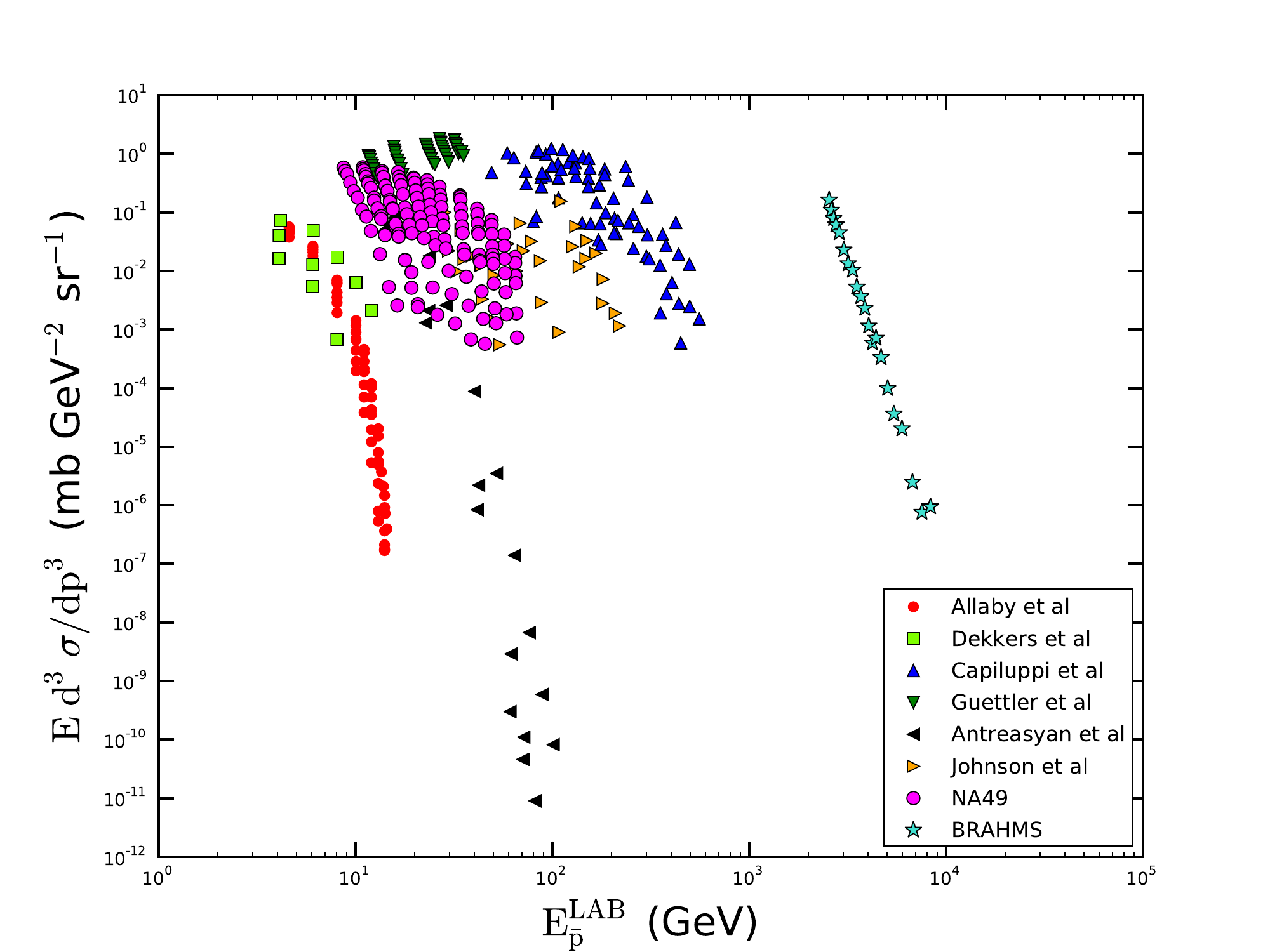}
	\end{subfigure}
	\hfill
	\begin{subfigure}[t]{0.495\textwidth}
		\centering
		\includegraphics[width=\linewidth]{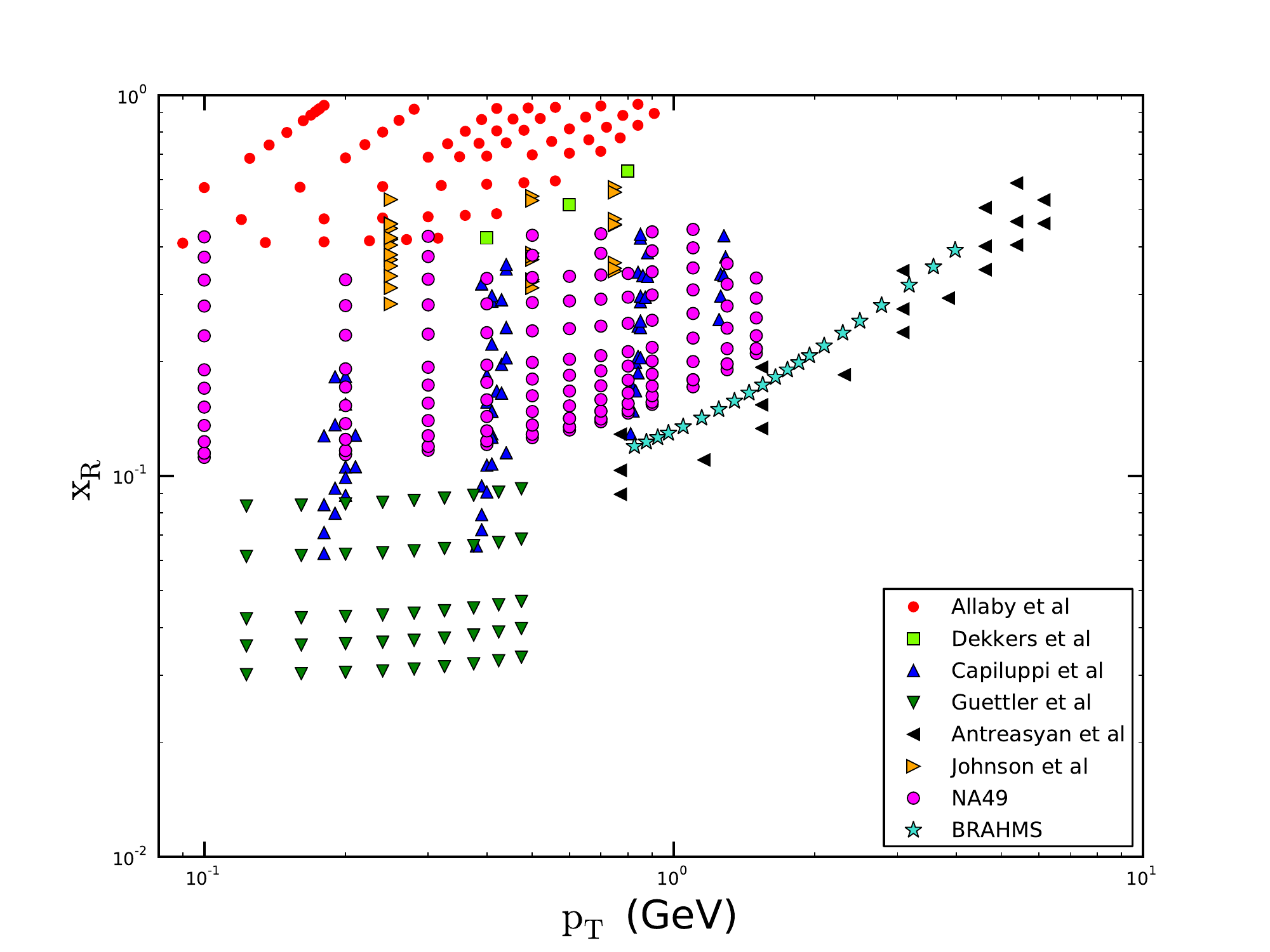}
	\end{subfigure}
	\caption{The data on $d^{3}\sigma_{pp\rightarrow\bar{p}} / dp^{3}$ as a function of $E_{\bar{p}}^{LAB}$ (left panel) and in the $p_{T}−x_{R}$ space (right panel)\cite{mauro2014}.}
	\label{graphic-illustration}
\end{figure}

Tab. \ref{datasets} summarizes datasets of antiproton production in $pp$ collisions along with their corresponding CM energy, $\sqrt{s}$, and $(p_{T}, x_{R})$ regions, where $p_{T}$ is the antiproton transverse momentum and $x_{R}$ is the radial-scaling variable given by\cite{mauro2014}
\begin{equation}
x_{R} = \frac{E^{*}_{\bar{p}}}{E^{*}_{\bar{p}.\text{max}}},
\end{equation}
where $E^{*}_{\bar{p}}$ is the antiproton energy and $E^{*}_{\bar{p}.\text{max}}$ is the maximal energy it can acquire; both are in the CM frame. Hence, the radial-scaling variable is always $\leq 1$. The maximal antiproton energy is given by\cite{mauro2014}:
\begin{equation}
\label{maximal_energy}
E^{*}_{\bar{p}.\text{max}} = \frac{s - 8 m_{p}^{2}}{2 \sqrt{s}}.
\end{equation}
It is to be noted that $E^{*}_{\bar{p}} \geq m_{p}$, which implies that $E^{*}_{\bar{p}.\text{max}} \geq m_{p}$; and since the square of the CM energy is $s = 2 m_{p}\left(E_{p} + m_{p}\right)$, where $E_{p}$ is the total energy of the incident proton in the LAB frame, it follows from \eqref{maximal_energy} that the threshold energy for the incident proton is $E_{p} \geq 7 m_{p}$. A graphic illustration of the same datasets is shown in Fig. \ref{graphic-illustration}. The differential cross section $d^{3}\sigma_{pp\rightarrow\bar{p}} / dp^{3}$ is shown as a function of $E_{\bar{p}}^{LAB}$, along with the corresponding combinations of $p_{T}$ and $x_{R}$\cite{mauro2014}. A key point to take into account when examining these datasets is that while the NA49 data was corrected for the effect of antiproton production from hyperon decays, other older datasets were not, which can be a significant source of systematic error.

\begin{table}[h]
\begin{center}
	\caption{Data sets of antiproton production in $pp$ collisions along with their corresponding $\sqrt{s}$ values and $(p_{T},x_{R})$ regions\cite{mauro2014}.}
	\label{datasets}
	\begin{tabular}{lccc}
		\hline\hline
		~ & ~ & ~ &\\
		Experiment & $\sqrt{s}$ (GeV) & $p_{T}$ (GeV) & $x_{R}$ \\
		\hline
		~ & ~ & ~ &\\
		Dekkers \textit{et al.}, CERN 1965\cite{PhysRev.137.B962} & $6.1, 6.7$ & $(0.00, 0.79)$ & $(0.34, 0.65)$ \\
		Allaby \textit{et al.}, CERN 1970\cite{Allaby:1970jt} & $6.15$ & $(0.05, 0.90)$ & $(0.40, 0.94)$ \\
		Capiluppi \textit{et al.}, CERN 1974\cite{CAPILUPPI1974189} & $23.3, 30.6, 44.6, 53.0, 62.7$ & $(0.18, 1.29)$ & $(0.06, 0.43)$ \\
		Guettler \textit{et al.}, CERN 1976\cite{GUETTLER197677} & $23.0, 31.0, 45.0, 53.0, 63.0$ & $(0.12, 0.47)$ & $(0.036, 0.092)$ \\
		Johnson \textit{et al.}, FNAL 1978\cite{PhysRevD.19.764} & $19.4, 23.8, 27.4$ & $(0.77, 6.15)$ & $(0.08, 0.58)$ \\
		Antreasyan \textit{et al.}, FNAL 1979\cite{GUETTLER197677} & $23.0, 31.0, 45.0, 53.0, 63.0$ & $(0.12, 0.47)$ & $(0.036, 0.092)$ \\
	BRAHMS, BNL 2008\cite{brahms2007} & $200$ & $(0.82, 3.97)$ & $(0.11, 0.39)$ \\
		NA49, CERN 2010\cite{na49-2010} & $17.3$ & $(0.10, 1.50)$ & $(0.11, 0.44)$ \\
		NA61, CERN 2017\cite{NA61}\footnote{Graphical illustration not included in Fig. \ref{graphic-illustration}.} & $6.3, 7.7, 8.8, 12.3, 17.3$ & --- & ---\\
		~ & ~ & ~ &\\
		\hline\hline
	\end{tabular}
\end{center}
\end{table}

The current focus is to cover the entire antiproton energy range of AMS-02, in order to be able to interpret the data and determine whether any exotic contributions do exist. That will require precise measurements of antiproton-production cross sections in $pp$ and $pHe$ collisions, covering the range from 1 GeV to 6 TeV for the proton-beam kinetic energy and the range from 2 to 8 for the pseudorapidity. As mentioned above, datasets of antiproton production in $pp$ collisions are very scarce, but data for all the other channels in Tab. \ref{tab:contributions} are almost non-existent. In fact, the first ever measurement of prompt antiproton-production in $pHe$ collisions was published by the LHCb collaboration only last August\cite{Aaij:2018svt}. The dataset was collected in fixed-target mode with a kinetic energy of 6.5 TeV for the proton beam, and covered a pseudorapidity range of $2 < \eta < 5$. The detected antiproton momenta ranged from 12 to 110 GeV/c. Another dataset was collected at LHCb in November 2016 but has not been published yet\cite{Denisov:2018unj}. On one hand, both sets of data are at very high energies; so will be all future expected datasets from the LHCb experiment.

On the other hand, the COMPASS collaboration is planning to perform a new fixed-target experiment at the M2 beam line of the CERN SPS to measure the antiproton-production cross sections of $pp$ and $pHe$ collisions\cite{Denisov:2018unj}. Using a proton beam of a few hundred GeV/c, this would be complementary to the LHCb measurements. This experiment can be carried out with proton-beam momenta ranging from a few tens of GeV/c up to 250 GeV/c; and would cover pseudorapidities $\eta > 2.4$. Meanwhile, the antiproton tracks would be identified and counted as a function of momentum and pseudorapidity. So far, it is planned to take measurements with proton-beam momenta of 50, 100, 190 GeV/c and the maximum momentum possible at CERN's M2 beam line.

\label{sec:pbar_production_NICA}
\section{\textsc{Antiproton Production at NICA}}~

The Nuclotron-based Ion Collider fAсility (NICA) is under construction at JINR\cite{Kekelidze:2017mhu}. The first priority program of NICA is the study of properties of dense baryonic matter in heavy ion collisions at the Multi-Purpose Detector (MPD)\cite{Toneev:2007yu}. In parallel, the operation of the NICA collider with polarized proton and deuteron beams for spin physics studies at the Spin Physics Detector (SPD) is also under consideration \cite{Savin:2014sva}. The planned kinetic energy of polarized proton beam is from 5 GeV to about 12.6 GeV while the polarized deuteron beam energy is expected to range from 4 GeV to 11.8 GeV. In $pp$ collisions at $\sqrt{s}=27$ GeV, the luminosity $L=10^{32}$ cm$^{-2}$ s$^{-1}$ should be achieved. Maximal luminosity of  $DD$ collisions is expected to be two orders of magnitude lower. Possibility to organize $pD$ collisions and operation with light ion beam like $^{3}$He is also assumed.

The SPD detector is considered as a universal $4\pi$ detector. It will consist of five main subsystems: i) vertex detector for precise reconstruction of the primary interaction point; ii) main tracker for track sign and momentum reconstruction in the magnetic field; iii) time-of-flight system for particle identification; iv) electromagnetic calorimeter for photon detection and electron and positron identification and v) outer detector for muon/hadron identification. A few configurations of magnetic systems based on solenoidal and toroidal fields and their combinations are under discussion but the final decision has not been made yet. The exact size and configuration of the SPD detector depends on this choice.

\begin{figure}[p]
	\centering
	\begin{subfigure}[t]{0.495\textwidth}
		\centering
		\includegraphics[width=\linewidth,height=0.3\textheight]{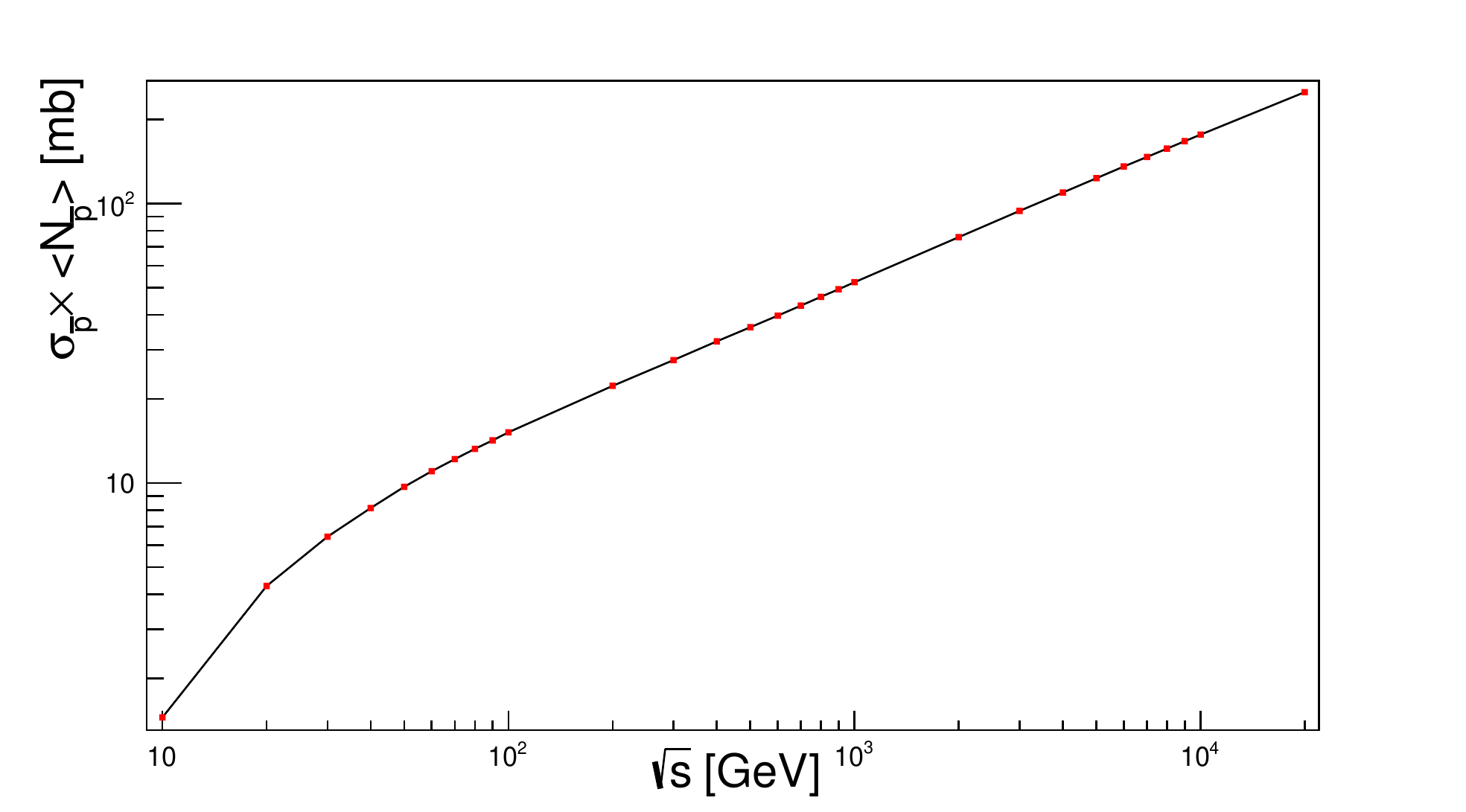}
		\caption{~}
		\label{sigma_pbar}
	\end{subfigure}
	\hfill
	\begin{subfigure}[t]{0.495\textwidth}
		\centering
		\includegraphics[width=\linewidth,height=0.3\textheight]{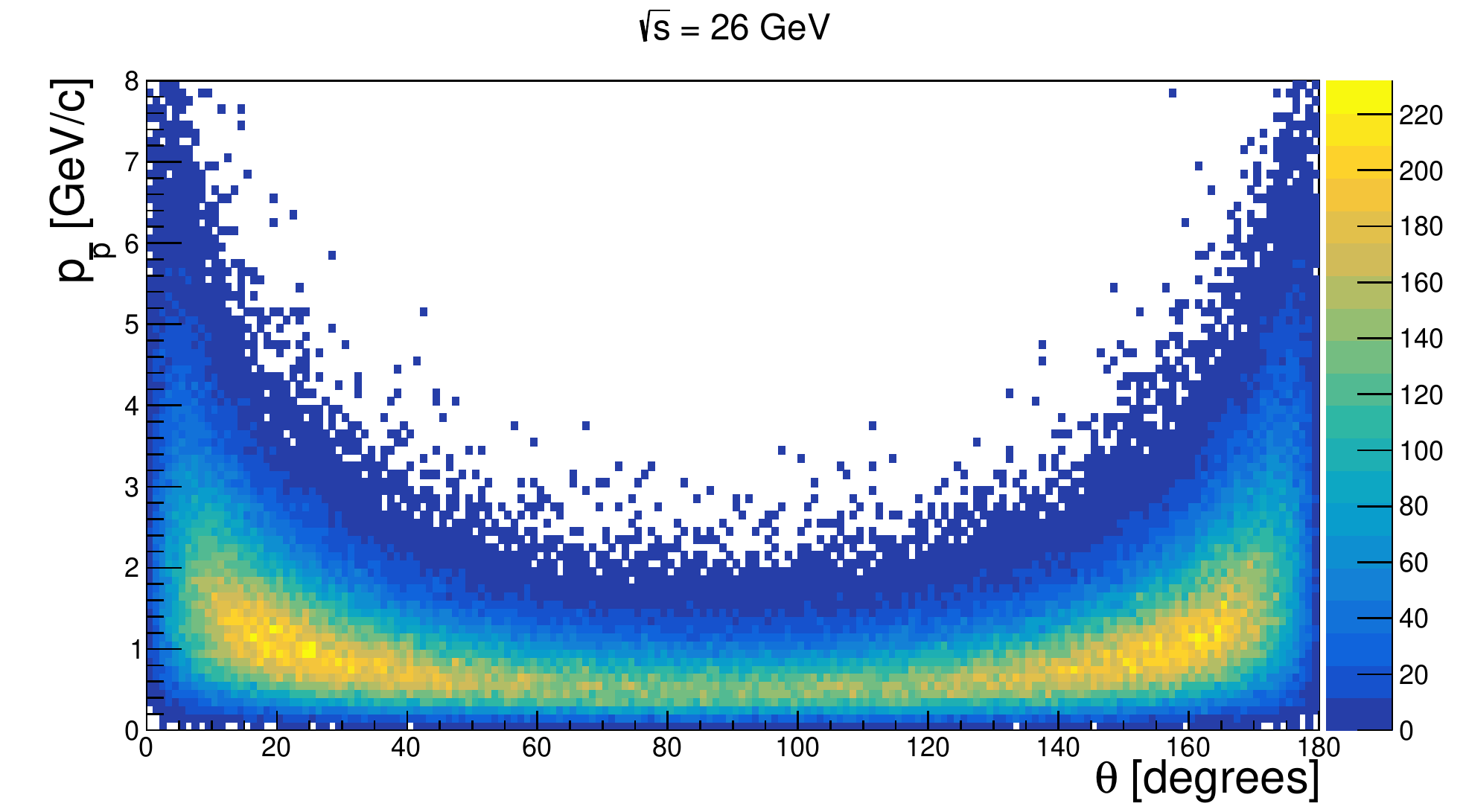}
		\caption{~}
		\label{angular_distribution}
	\end{subfigure}
	\caption{\label{fig:spectra}(a) The antiproton-production cross section in $pp$ collisions multiplied by the average antiproton-multiplicity versus the CM energy; covering a range from 10 GeV to 20 TeV. (b) The transverse momentum ($p_{T}$) versus the polar angle ($\theta$) of antiprotons produced in $pp$ collisions at $\sqrt{s} = 26$ GeV.}
\end{figure}

\begin{figure}[p]
	\centering
	\begin{subfigure}[t]{0.495\textwidth}
		\centering
		\includegraphics[width=\textwidth,height=0.3\textheight]{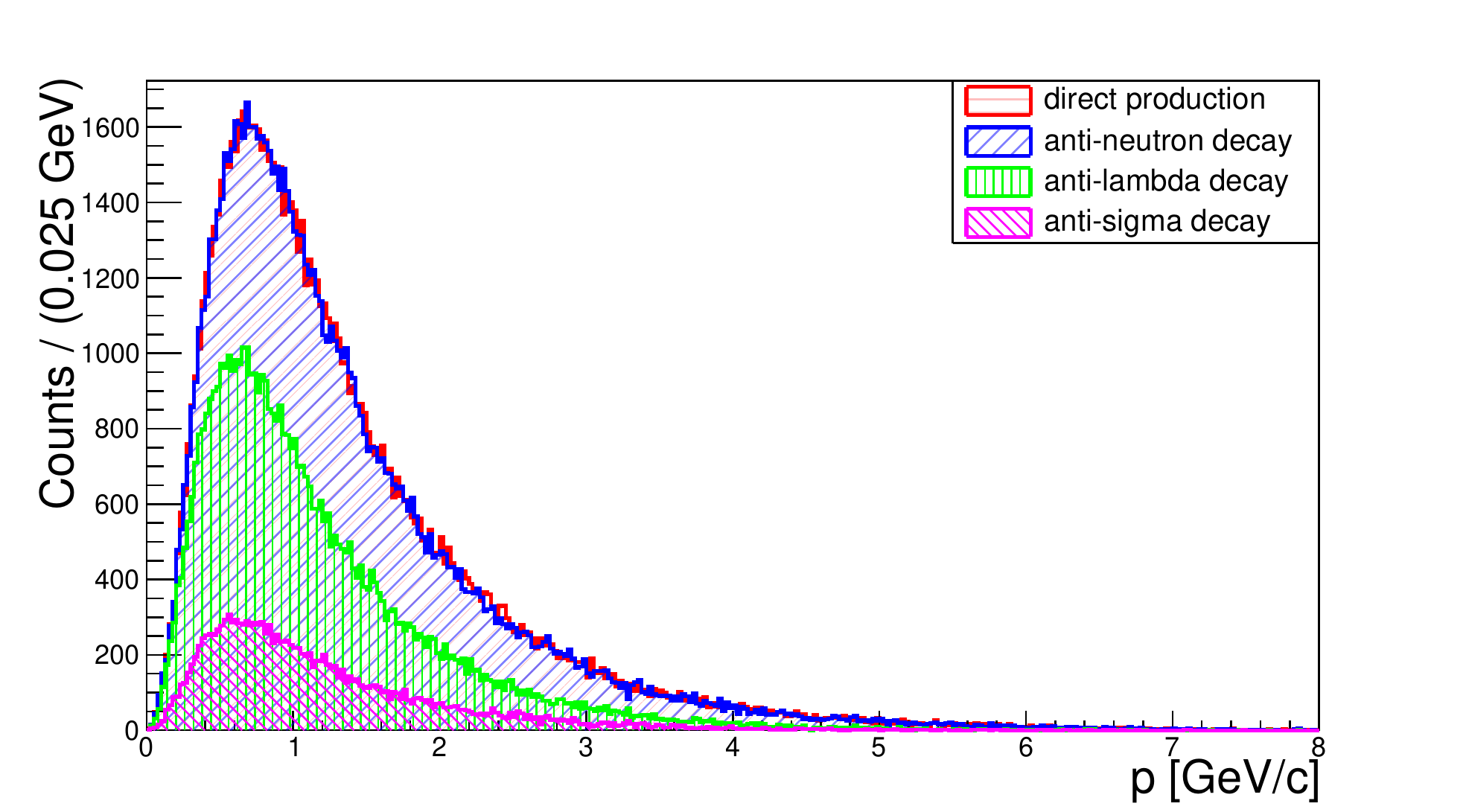}
		\caption{~}
		\label{energy_spectra}
	\end{subfigure}
	\hfill
	\begin{subfigure}[t]{0.495\textwidth}
		\centering
		\includegraphics[width=\textwidth,height=0.3\textheight]{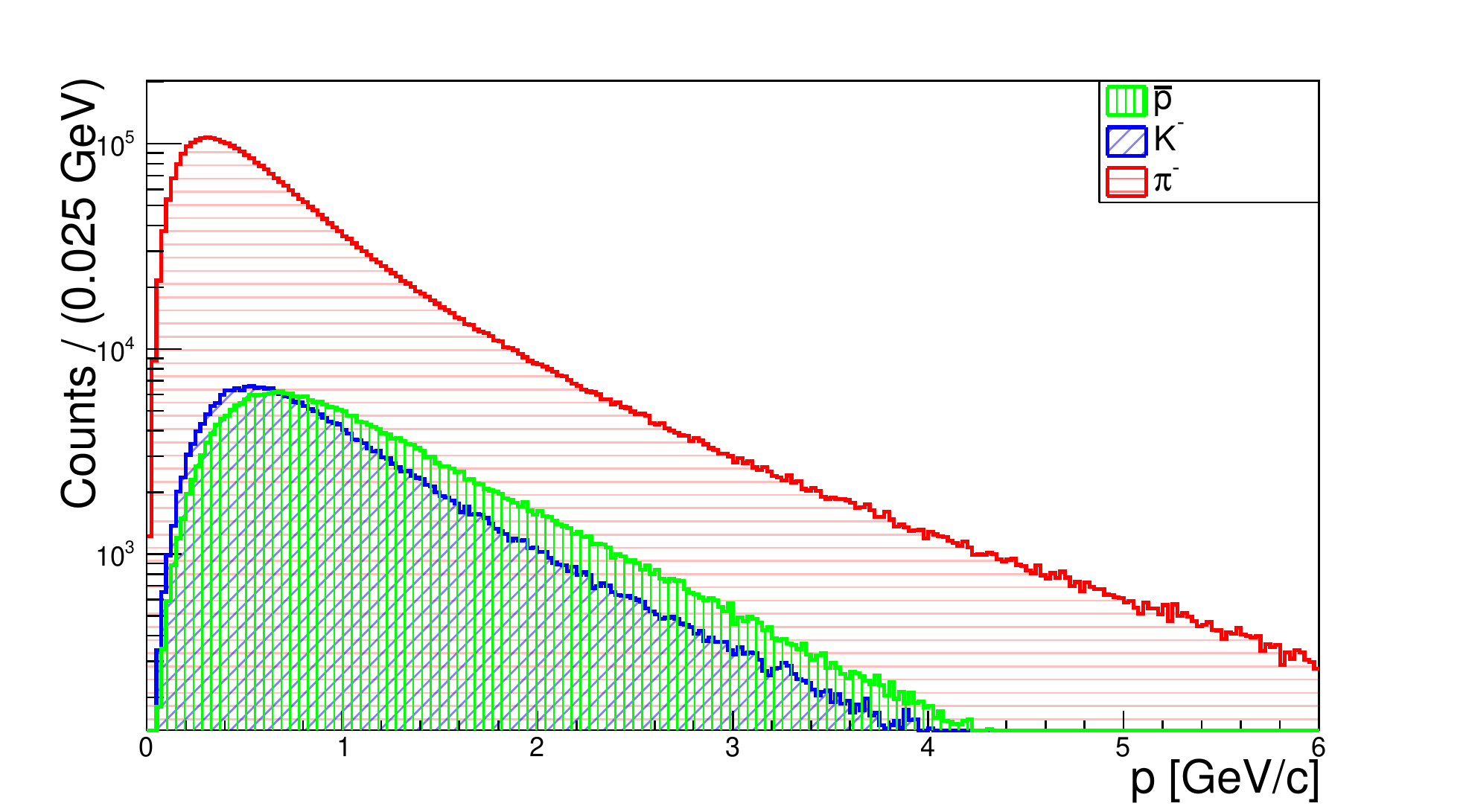}
		\caption{~}
		\label{particles_momenta}
	\end{subfigure}
	\caption{The momentum spectrum of (a) the different antiproton-contributions, and (b) total antiprotons, and negatively charged kaons ($K^{-}$) and pions ($\pi^{-}$), produced in $pp$ collisions at CM energy $\sqrt{s} = 26$ GeV.}
\end{figure}

\begin{figure}[!htb]
	\centering
	\begin{subfigure}[t]{0.495\textwidth}
		\centering
		\includegraphics[width=\linewidth,height=0.3\textheight]{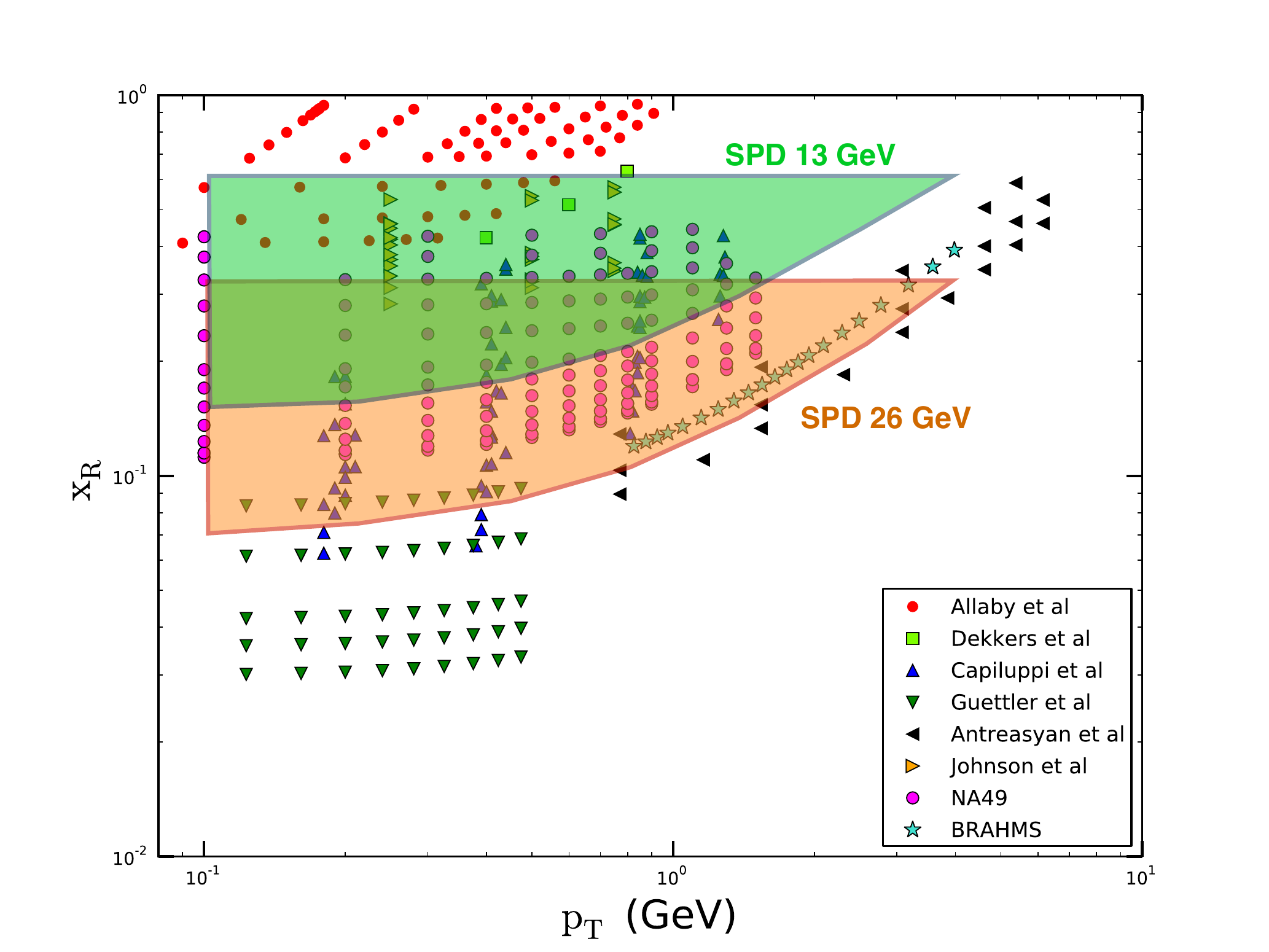}
		\caption{~}
	\end{subfigure}
	\hfill
	\begin{subfigure}[t]{0.495\textwidth}
		\centering
		\includegraphics[width=\linewidth,height=0.285\textheight]{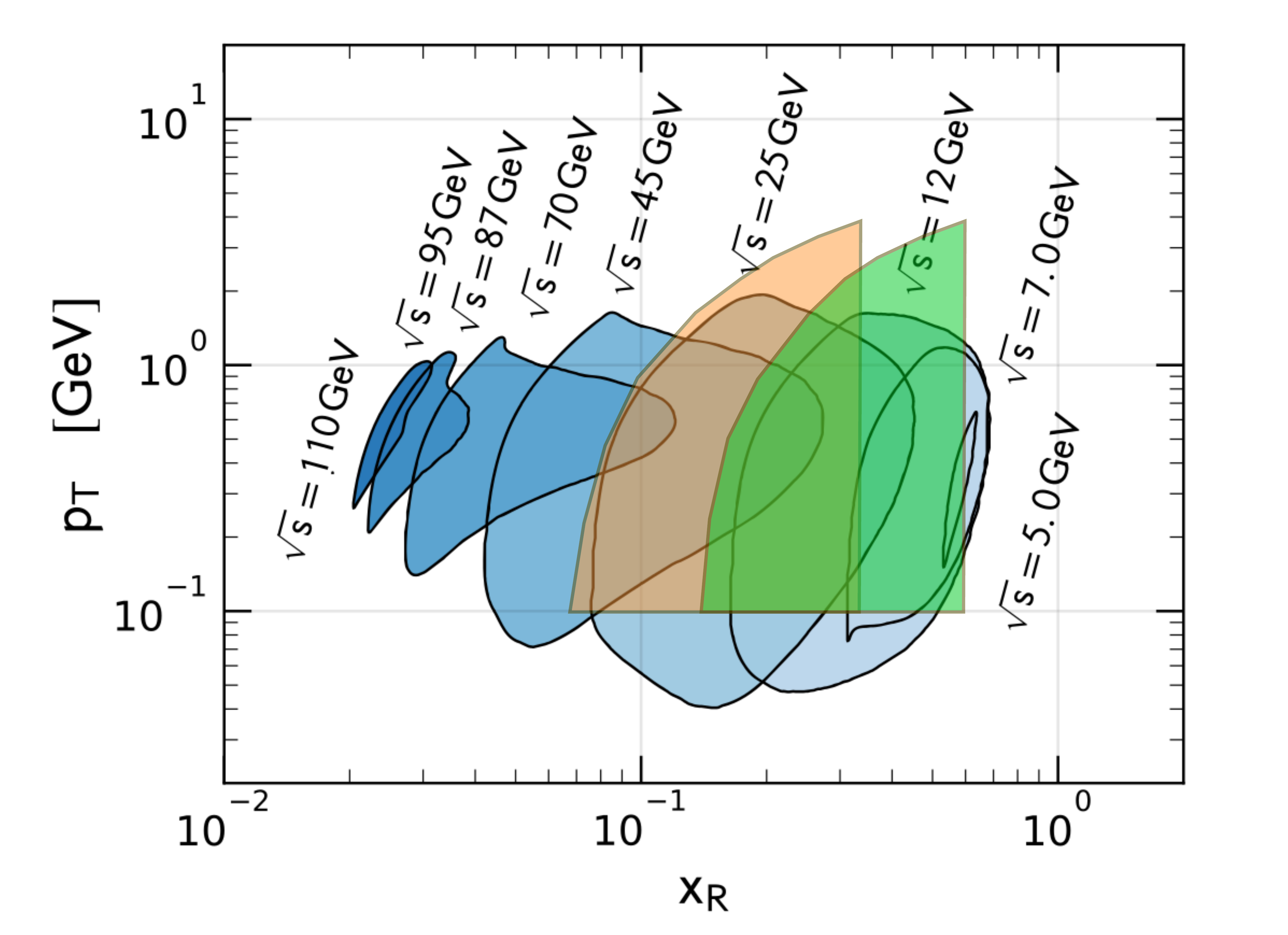}
		\caption{~}
	\end{subfigure}
	\caption{The kinematic range accessible by by $pp$ collisions at SPD detector, superimposed on \mbox{(a) the} graphic illustration of currently existing data\cite{mauro2014}, and \mbox{(b) required} measurement range to match the uncertainty level of AMS-02 measurements\cite{PhysRevD.96.043007}. Some cuts were applied to the momenta ($p$), the transverse momenta ($p_{T}$), and the polar angle of the produced antiprotons, based on the current conceptual design of the SPD.}
	\label{NICA_range}
\end{figure}

\begin{table}[!htb]
	\begin{center}
		\caption{The relative yields of antiprotons in $pp$ collisions at CM energies $\sqrt{s} = 13$ and 26 GeV.}
		\label{tab:relative_yields}
		\begin{tabular}{ccc}
			\hline\hline
			Production & $ \sqrt{s} = 13 $ GeV & $ \sqrt{s} = 26 $ GeV \\
			\hline
			Direct         &  38.6 &  37.3\\
			$\bar{n}$ decay         &   37.9  & 36.9\\
			$\bar{\Lambda}$ decay    &  18.1    & 19.8\\
			$\bar{\Sigma}^{-}$ decay    &  5.4  & 6.0 \\
			\hline\hline
		\end{tabular}
	\end{center}
\end{table}

The SPD detector with its $4\pi$ acceptance and particle identification capabilities looks promising for study of antiproton production in $pp$ and $DD$ collisions in the corresponding energy range, where Fig. \ref{sigma_pbar} shows the expected yield of antiprotons in $pp$ collisions in a wide energy-range, based on Monte Carlo simulation generated with Pythia8---the same as all distributions included in this Section, while Fig. \ref{angular_distribution} shows the spectra of the transverse momenta ($p_{T}$) of produced antiprotons as a function of the polar angle ($\theta$). As one can see from Fig. \ref{energy_spectra}, the momentum of antiprotons produced in proton-proton collisions is low enough for their identification by the time-of-flight method. Assuming time resolution of about 100 ps and of about 2 meters of the flight base, we can conclude that $K^{-}/\bar{p}$ separation would be possible for protons with momentum up to $4$ GeV/c, which would cover the bulk of the yields, as can be seen from Fig. \ref{particles_momenta}. Magnetic field with typical intensity $0.5$ T defines the minimal detectable $p_{T}$ of produced antiprotons to be about $0.1$ GeV/c. Material budget in the central part of the spectrometer could also limit the minimal momentum of detected antiprotons. More detailed estimations could be made after full Monte Carlo simulation of the SPD setup. The kinematic range that could be covered by $pp$ collisions at SPD detector, in comparison with both the currently-existing data and the required coverage for matching the AMS-02 data, is presented in Fig. \ref{NICA_range}. Moreover, the ability to reconstruct secondary vertices could make it easy to investigate the yield of antiprotons from decays of hyperons; where Tab. \ref{tab:relative_yields} shows the significance of the contribution of hyperon decay to the total antiproton production, and how it increases with energy. In addition, the capability of the electromagnetic calorimeter to be used for detection of low-energy neutrons should be also investigated.

\label{sec:conclusions}
\section{\textsc{Conclusion}}~

The Spin Physics Detector planned as a universal multipurpose setup at the NICA collider for comprehensive study of proton and deuteron interactions could make a sizable contribution to the search of physics beyond the Standard Model. We propose to use it for the precision measurements of the differential cross section of antiproton production in $pp$ and $pD$ collisions required by astrophysical searches for dark matter. The SPD can measure energy and angular distributions of antiprotons produced both directly and from the decays of $\bar{\Lambda}$ and $\bar{\Sigma}^{-}$ hyperons in the kinematic range starting from the threshold. The collider mode and the $4\pi$ geometry of the SPD detector provide a unique possibility to study the production of antiprotons at high transverse momenta (up to $p_T \sim \sqrt{s}/2$) which is unavailable for fixed-target experiments. Possibility for NICA to operate with beams of light nuclei, like $^{3}$He and $^4$He, would extend this programme. The main requirements for the SPD setup in order to perform the measurements are an advanced particle identification (e.g. Time-of-Flight system), low material budget, precision reconstruction of secondary vertices, and the possibility to perform absolute luminosity measurement with accuracy of about 1\%. Since at the moment only a general concept of the SPD setup exists, it is not possible to discuss the accuracy of proposed measurements. More detailed Monte Carlo studies are needed for that.

\bibliographystyle{pepan}
\bibliography{ref}

\end{document}